\documentclass[pdflatex,prd,twocolumn,showpacs,superscriptaddress,nofootinbib]{revtex4-2}

\usepackage{amssymb,amsmath,color,graphicx,multirow,tabularx,physics,comment}
\usepackage[normalem]{ulem}
\usepackage[hyperfootnotes=false]{hyperref}
\hypersetup{
    colorlinks=true,
    linkcolor=blue,
    citecolor=blue,
    urlcolor=blue,
    linktoc=page
}

\usepackage{davitex}
\usepackage[dvipsnames]{xcolor}

\newcommand \hmu {\hat{\mu}}
\newcommand \hchi {\hat{\chi}}

\usepackage{}

\newcommand \eq[1] {Eq.\ (\ref{#1})}
\newcommand \Tpco {T_{\rm pc,0}}

\newcommand \beq {\begin{equation}}
\newcommand \eeq {\end{equation}}
\newcommand \beqa {\begin{eqnarray}}
\newcommand \eeqa {\end{eqnarray}}

\newcommand \calZ {\mathcal{Z}}

\begin{document}

\title{A generalized definition of the isothermal compressibility in (2+1)-flavor QCD} 

\author{D. A. Clarke}
\affiliation{Department of Physics and Astronomy, University of Utah, Salt Lake City, Utah, United States}
\author{J. Goswami}
\author{F. Karsch}
\affiliation{Fakult\"at f\"ur Physik, Universit\"at Bielefeld, Bielefeld, Germany}
\author{P. Petreczky}
\affiliation{Physics Department, Brookhaven National Laboratory, Upton, New York, United States}

\date{\today}

\begin{abstract}
We introduce a generalized definition of the isothermal compressibility ($\kappa_{T,\sigma_Q^2}$) calculable by keeping net conserved charge fluctuations 
rather than total number densities constant.
We present lattice QCD results for this isothermal compressibility, expressed in terms of fluctuations of conserved charges that are related to baryon ($B$), electric charge ($Q$) and strangeness ($S$) quantum numbers.
This generalized isothermal compressibility is compared with hadron resonance gas model
calculations as well as with heavy-ion collision data obtained at RHIC and the LHC.
We find $\kappa_{T,\sigma_Q^2}=13.8(1.3)$~fm$^3$/GeV at $\Tpco=156.5(1.5)$~MeV and $\hmu_B=0$. This finding
is consistent with the rescaled result of the ALICE Collaboration, where we replaced the number of charged hadrons ($N_{\rm ch}$) by the total number of hadrons ($N_{\rm tot}$) at freeze-out.  Normalizing this result with the QCD pressure ($P$) we find that the isothermal compressibility on the pseudo-critical line stays close to that of an {\it ideal gas}, {\it i.e.} $P \kappa_{T,\sigma_Q^2}\simeq 1$.
\end{abstract}


\maketitle

{\it Introduction ---}
A central goal of theoretical and experimental studies of strong-interaction
matter is to understand its different phases and the properties of matter in these
phases. Thermal properties of strong-interaction matter \cite{Rajagopal:2000wf,Busza:2018rrf}, e.g. the equation of state, velocity of sound or its viscosity, are widely studied experimentally in heavy-ion collisions 
(HIC) and theoretically in the framework of the theory of strong interactions, Quantum Chromodynamics (QCD). The small shear and bulk viscosities, determined
in HIC and calculated in lattice QCD simulations \cite{Meyer:2007ic,Dusling:2011fd,Altenkort:2022yhb}, led to the conclusion 
that this matter is an ``almost perfect fluid" \cite{Heinz:2005zg}. 
We will show here that in the vicinity of the pseudo-critical temperature, which corresponds
to the freeze-out temperature in HIC, the isothermal compressibility stays close to that of an ideal gas.

The current phenomenological
discussion \cite{Mrowczynski:1997kz,Hartnack:2005tr,Dexheimer:2007mt}
and experimental determination \cite{PHENIX:2008psu,Mukherjee:2017elm,ALICE:2021hkc} 
of the isothermal compressibility ($\kappa_T$) of strong-interaction matter relies on 
a statistical physics definition
\cite{Mrowczynski:1997kz,Landau_1958,Huang_1987}, which is valid for systems in which particle
number can be considered as a well defined conserved quantity,
\begin{equation}
  \kappa_T \;=\; -\frac{1}{V}\left(\frac{\partial V}{\partial P}\right)_{T,N}
  = \frac{1}{n^2}\frac{\partial^2 P}{\partial \mu^2}=\frac{1}{T^4\,\hat n^2}\,
\frac{\partial^2\hat P}{\partial\hat\mu^2}
  \, ,
  \label{eq:kappa}
\end{equation}
with pressure $P$, spatial volume $V$, temperature $T$,
$n=N/V$ denoting the particle number density, and $\mu$ being the 
chemical potential coupling to the number operator.
Here and in the following we use the notation
$\hat{O}\equiv OT^{-k}$
with $k\in\Z$ chosen so that $\hat{O}$ is dimensionless. The application of such an approach to the analysis of properties 
of strong-interaction matter,
created in HIC at the time of hadronization (freeze-out),
makes use of the fact that relativistic hadron resonance
gas (HRG) models provide a rather good description of the thermodynamics
of QCD in the vicinity of the freeze-out temperature. 

The concept of well defined particle number densities, however, cannot be carried over
straightforwardly to a first-principle, QCD-based calculation of the 
isothermal compressibility of matter, where the conserved charges are related to differences of particle and anti-particle numbers rather
than total particle numbers, {\it i.e.} net-baryon number ($N_B$), net-electric charge ($N_Q$) 
and net-strangeness ($N_S$). In fact, defining $\kappa_T$ by keeping net-number densities rather than total number
densities fixed amounts to replacing in \eq{eq:kappa} the total number
density $n$ by a net-charge number density $n_Q$ or $n_B$. 
In the limit of vanishing chemical potential this clearly
leads to a divergent isothermal compressibility which easily is verified in 
non-interacting HRG model calculations and explicitly shown in lattice QCD
calculations \cite{Gupta:2014qka}.

In this letter we will discuss a generalized version of the
isothermal compressibility which does not explicitly rely on the particle 
content of a thermal medium but makes use of conserved charges and their 
fluctuations that are accessible in QCD calculations as well
as measurements in HIC. We consider the isothermal 
compressibility of strong-interaction matter at fixed value of the
ratio of net-electric charge to net-baryon number, $r_Q=N_Q/N_B$, 
and fixed net-strangeness to net-baryon number ratio, $r_S=N_S/N_B$.
Furthermore, the partial derivative of volume with respect to pressure
is taken, keeping a function ($f_Q$) of the net-electric charge fixed,
\begin{equation}
  \kappa_{_{\scriptstyle T,f_Q,r_Q,r_S}} \;=\; -\frac{1}{V}\left(\frac{\partial V}{\partial P}\right)_{T,f_Q,r_Q,r_S}
  \, .
  \label{eq:kappafQ}
\end{equation}
In particular we consider the cases of fixed net-electric charge, 
$f_Q=V \partial P/\partial \mu_Q\equiv N_Q$, and fixed net-electric charge fluctuations, 
$f_Q=VT \partial^2 P/\partial \mu_Q^2\equiv \sigma^2_Q$. The latter choice is motivated by
the observation that 
total particle numbers in a non-interacting HRG are well approximated by net-charge fluctuations
and that HRG models provide a good description of lattice QCD calculations close
to the freeze-out temperature.

However, as pointed out, the former choice leads to a divergent compressibility at vanishing chemical potentials even in HRG model calculations.

{\it Isothermal compressibility in HRG models ---}
We briefly recapitulate here the calculation of the isothermal compressibility $\kappa_{T,\vec{N}}$ in non-interacting HRG
models, which is performed for a fixed number of particles and anti-particles ($\vec{N}$)
to circumvent the problem of divergences at $\mu=0$, and which is used also in experimental determinations 
of $\kappa_{T,\vec{N}}$. Calculations of $\kappa_{T,\vec{N}}$ in
the framework of HRG models take advantage of the
fact that in a grand canonical prescription of the HRG model the pressure and other thermodynamic observables are explicitly given in terms of sums of contributions from different particle species. 
Although the grand canonical ensemble generally is defined as a function of $T$, $V$, and conserved‑charge chemical potentials
$\vec{\mu}=(\mu_B, \mu_Q, \mu_S)$,  
\begin{align}
P &= \lim_{V\to\infty}\frac{T}{V}\,\ln \calZ\!\bigl(T,V,\vec{\mu}\bigr)\; ,
\end{align}
in a non-interacting HRG model, one may project onto different particle and anti-particle sectors by introducing
a chemical potential, $\mu_i$, for each particle species $i$, which we set to zero after taking appropriate derivatives, {\it i.e.}
we introduce $\mu_{H,i}=B\mu_B+Q\mu_Q+S \mu_S+\mu_i$ for each hadron with quantum numbers $(B,Q,S)$.
With this one obtains particle numbers and
fluctuations in each sector \cite{Mrowczynski:1997kz,Mukherjee:2017elm}
\begin{align}
N_i &= \left. V \frac{\partial P}{\partial \mu_i}\right|_{\mu_i=0}=Vn_i=VT^3\hat{n}_i \; ,\label{eq:N}\\[2pt]
\sigma_{i}^2 &=\left. T \frac{\partial N_i}{\partial \mu_i}\right|_{\mu_i=0} = VT \frac{\partial^{2} P}{\partial \mu_i^{2}}=VT^3\hchi_2^i \; .
\label{eq:sigma}
\end{align}

The inverse of the isothermal compressibility in a non-interacting HRG is then obtained as \cite{Mukherjee:2017elm}
\begin{align}
    \kappa_{T,\vec{N}}^{-1} = -V \left( \frac{\partial P}{\partial V}\right)_{T,\vec{N}}
= T^4\hspace*{-3mm}\sum_{i\in {\rm hadrons}} \frac{\hat{n}_i}{\omega_{i}} \; ,
\label{eq:kinv}
\end{align}
where $\omega_i=\sigma_{i}^2/N_i$ are the scaled variances.

In a non-interacting HRG, Boltzmann-statistics is a good approximation for all baryons. As a consequence,
for all baryons $\omega_i=1$  for all $T$ and $\vec{\mu}$. For mesons appreciable deviations
from unity only arise for pions and kaons. In all other cases $\omega_i\simeq 1$ to better than 0.2\%.  

In order to make contact with experimental determinations of $\kappa_{T,\vec{N}}$, which is based on a measurement of charged particle yields and fluctuations, we separate in \eq{eq:kinv} contributions from charged and neutral 
hadrons,
\begin{align}
    \kappa_{T,\vec{N}}^{-1} = T^4\hspace*{-3mm}\sum_{i\in {\rm charged}} \frac{\hat{n}_i}{\omega_{i}} + T^4\hspace*{-3mm}\sum_{i\in {\rm neutral}} \frac{\hat{n}_i}{\omega_{i}}. 
    \label{eq:kinv-exact} 
\end{align}
As in HRG model calculations the scaled variances for different particle species vary little, introducing
a common scaled variance for all charged ($\omega_{\rm ch}$), neutral ($\omega_{\rm 0}$), or even all hadrons ($\omega_{\rm tot}$) provides a good approximation,
    \begin{align}
  \kappa_{T,\vec{N}}^{-1}   
&\simeq T^4 \frac{\hat{n}_{\rm ch}}{\omega_{\rm ch}}+T^4 \frac{\hat{n}_{0}}{\omega_{0}} 
\simeq T \frac{n_{\rm tot}}{\omega_{\rm tot}}\; ,
\label{eq:kinv-approx}
\end{align}
where we introduced
\begin{align}
\hat{n}_X =\sum_{i\in X} \hat{n}_i \;\; ,\;\;
\omega_X =\frac{\sum\limits_{i\in X} \sigma^2_i}{\sum\limits_{{i}\in X} N_i} \;\; ,\;\;
X= {\rm ch},0, {\rm tot} 
\end{align}
for charged, neutral and total number of hadrons, respectively.

We note that the determination of $\kappa_{T,\vec{N}}$ from HIC on particle
yields and fluctuations
\cite{PHENIX:2008psu,Mukherjee:2017elm,ALICE:2021hkc} is based 
only on the first term 
in the middle expression
of \eq{eq:kinv-approx}. I.e. the contribution of
neutral hadrons to the compressibility is neglected.
To the extent that the total number of hadrons in HRG model calculations as well as in heavy-ion collisions
is almost a factor of two larger than the number of charged hadrons 
\cite{Braun-Munzinger:2014lba,ALICE:2021hkc}
this would reduce $\kappa_{T,\vec{N}}$ by almost a factor 2.

In Table~\ref{tab:kappaHRG} we give values for $\hat{n}_X$ and $\omega_X$ at the pseudo-critical temperature, $\Tpco = 156.5$~MeV and $\mu_B=0$  for a 
non-interacting HRG\footnote{All HRG model calculations shown in this work are obtained by using the QMHRG2020 hadron list \cite{Bollweg:2021vqf}.}. Using also the QMHRG2020 result for the pressure, $P=0.7630\;\Tpco^4$, we thus obtain 
\begin{align}
    \kappa_{T,\vec{N}}T^4 =1.3441 \;\; ,\;\; P \kappa_{T,\vec{N}} = 1.0256 \; ,
    \label{eq:kappa-p}
\end{align}
which is close to the isothermal compressibility of a medium obeying the ideal gas equation of state (EoS)\footnote{In non-interacting HRG models, significant deviations from the ideal gas EoS only
arise from pions and kaons, which need to be treated as Bose gases. All other hadrons can be treated
in Boltzmann approximation for which a ideal gas EoS is a good approximation.}
($PV=NkT$), {\it i.e.} $\kappa_{T,N}=1/P$.

\begin{table}[htbp]
  \centering
  \caption{Summary of charged and neutral particle multiplicities and scaled variance in a non-interacting HRG at $T=156.5$~MeV.} 
\label{tab:kappaHRG}
  \begin{ruledtabular}
  \begin{tabular}{lcc}
    \textrm{Hadrons} & $\hat{n}_{X}$ & $\omega_{X}$ \\
    \hline
    \multicolumn{3}{l}{\textbf{Mesons}} \\
    \quad Charged [X=ch]     &  0.3448 &  1.0667 
\\
    \quad Neutral [X=0]      &  0.3059             & 1.0418         \\[1mm]
    \multicolumn{3}{l}{\textbf{Baryons}} \\
    \quad Charged [X=ch]     & 0.0726             & 1.0          \\
    \quad Neutral [X=0]      &  0.0549             & 1.0         \\     [1mm]
    \multicolumn{3}{l}{\textbf{Combined}} \\
    \quad Charged [X=ch]     & 0.4174             & 1.0551          \\
    \quad Neutral [X=0]      & 0.3608             & 1.0354         \\
    \hline
    \textbf{All hadrons} &  0.7782 & 1.0460
  \end{tabular}
  \end{ruledtabular}
\end{table}
{\it Generalized isothermal compressibility ---}
The HRG model calculation discussed above of course
has the disadvantage that it explicitly relies on 
the introduction of point-like, non-interacting particles as fundamental (conserved charge) degrees of freedom. 
That is not possible in QCD, and even in the HRG context, it
ignores that the truly conserved charges are $B, Q, S$
and eventually also the quantum numbers of heavier quarks.
Rather than defining the isothermal compressibility
by using $\partial V/\partial P$ taken at fixed $(T,\vec{N})$
we thus introduce a generalized version of the
isothermal compressibility for strangeness-neutral ($r_S=0$) (2+1)-flavor QCD
with fixed $r_Q$,

\begin{align}
    \kappa_{_{\scriptstyle T,f_Q,r_Q,0}}T^4 = -\frac{1}{f_Q \hat{n}_B}\left(\frac{\partial f_Q}{\partial \hmu_B}k_B+\frac{\partial f_Q}{\partial \hmu_Q}k_Q+\frac{\partial f_Q}{\partial \hmu_S}k_S\right)\; ,
\label{eq:kappars0}
\end{align}
where we keep a function $f_Q$ of the net-electric charge $Q$ fixed. 
The partial derivatives on the right hand side are understood to be taken at fixed $T,V$ as well as 
keeping the remaining two chemical potentials fixed, and
\begin{align}
k_B &=  - (1+(2r_Q-1) A)/D
\; ,
        \\[2mm]
k_S &=  (\hchi_{11}^{BS}/\hchi_2^{S}-(2r_Q-1) A)/D
\; ,
\\[2mm]
k_Q &= 2(2r_Q-1) A/D
\; ,
\end{align}
with $D=1-(2r_Q-1)^2 A$, and
\begin{align}
     A= \frac{ \hchi_2^S \hchi_{11}^{BQ} -\hchi_{11}^{BS}
    \hchi_{11}^{QS}}{\hchi_2^{S} (\hchi_{11}^{BQ}-2
    \hchi_2^{Q}+\hchi_{11}^{QS})} \; .
\end{align}
Here the dimensionless generalized susceptibilities \cite{Bazavov:2017dus} are defined as 
\begin{equation} 
  \hchi^{BQS}_{ijk}(T,V,\vec{\mu})
  =
    \frac{1}{VT^3}\,
    \frac{\partial^{\,i+j+k}\ln\mathcal{Z}(T,V,\vec\mu)}
         {\partial\hat\mu_B^i\,\partial\hat\mu_Q^j\,\partial\hat\mu_S^k}
    \;.
  \label{eq:suscept}
\end{equation}
Using results for second-order cumulants, given in \cite{Bollweg:2021vqf}, we find $A=-0.071(2)$ in QCD and $A=-0.070$ in HRG.
Differences in the isothermal compressibility obtained in the isospin symmetric limit, $r_Q=1/2$, and in the case of interest in HIC experiments, $r_Q=0.4$, respectively, thus are smaller than 1.5\%. In the following we 
therefore discuss only the isospin symmetric case $r_Q=1/2$.
In this case \eq{eq:kappars0} reduces to\footnote{In the following we suppress the subscripts for
$r_Q=1/2$ and $r_S=0$.}

\begin{align}
    \kappa_{_{\scriptstyle T,f_Q}}T^4 = \frac{1}{f_Q \hat{n}_B}\left( \frac{\partial f_Q}{\partial \hmu_B} - \frac{\hchi_{11}^{BS}}{\hchi_2^S} \frac{\partial f_Q}{\partial \hmu_S}\right)\; . 
\end{align}
We discuss the cases of fixed net-electric charge\footnote{Lattice QCD results for the isothermal compressibility at fixed net-baryon number have been presented in \cite{Gupta:2014qka}. 
They clearly show the strong divergence appearing in $\kappa_{T,N_Q}$ as well as
$\kappa_{T,N_B}$ in the limit of vanishing chemical potentials.}, $f_Q=N_Q$, and
fixed electric charge fluctuations, 
$f_Q=\sigma_Q^2=VT^3\hchi_2^Q$. For these two cases we find 

\vspace*{2mm}

\noindent
\underline{$f_Q=N_Q$:}
\begin{align}
            \kappa_{_{\scriptstyle T,N_Q}} T^4 
    &= \frac{\hchi_{11}^{BQ}}{\hat{n}_B \hat{n}_Q}\left( 1- \frac{\hchi_{11}^{QS}\hchi_{11}^{BS}}{\hchi_{11}^{BQ}\hchi_2^S}   \right)\; ,
\label{eq:mod-kQ1}
\end{align}
\noindent
\underline{$f_Q=\sigma_Q^2=VT^3\hchi_2^Q$:}
\begin{eqnarray}
            \kappa_{_{\scriptstyle T,\sigma_Q^2}} T^4 
    &=& \frac{\hchi_{12}^{BQ}}{\hat{n}_B\hchi_2^Q}\left( 1- \frac{\hchi_{21}^{QS}\hchi_{11}^{BS}}{\hchi_{12}^{BQ}\hchi_2^S}   \right)  \nonumber \\
    &=& \kappa_T^{(0)} \left(1 + \delta\kappa_T\right)T^4  \;\; .
\label{eq:mod-kQ2}
\end{eqnarray}
We note that $\kappa_{T,N_Q}\sim \hat{n}_Q^{-2}$ diverges in the limit of vanishing chemical
potentials, while 
the appearance of ratios of first- and third-order cumulants in  \eq{eq:mod-kQ2} ensures that 
the isothermal compressibility at fixed $f_Q=VT^3\hchi_2^Q$ has a finite limit for
vanishing chemical potentials.

\begin{figure}
\centering
\includegraphics[width=0.92\linewidth]{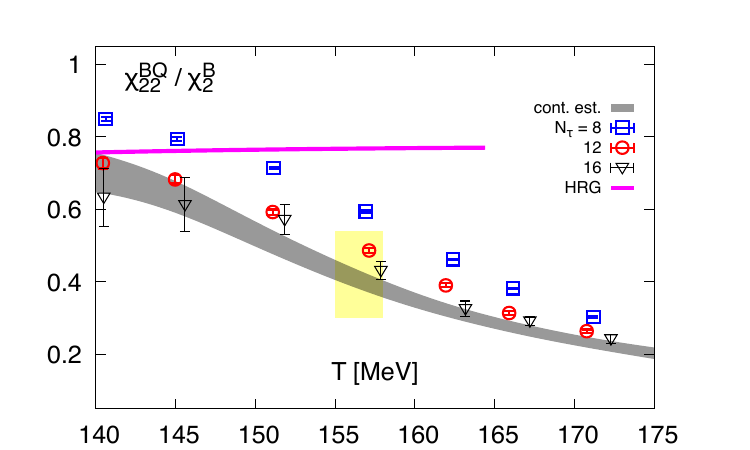}
\caption{Ratio of fourth-order cumulant of electric charge and baryon number correlations and 
the second-order cumulant of baryon number fluctuations for $\vec{\mu}=0$ 
versus $T$. Shown are results from (2+1)-flavor QCD and HRG model
calculations. The grey band shows the weighted average of continuum limit extrapolations 
obtained with and without including the $N_\tau=8$ data set.}
\label{fig:BQ22}
\end{figure}

\begin{figure*}
\centering
\includegraphics[width=0.43\linewidth]{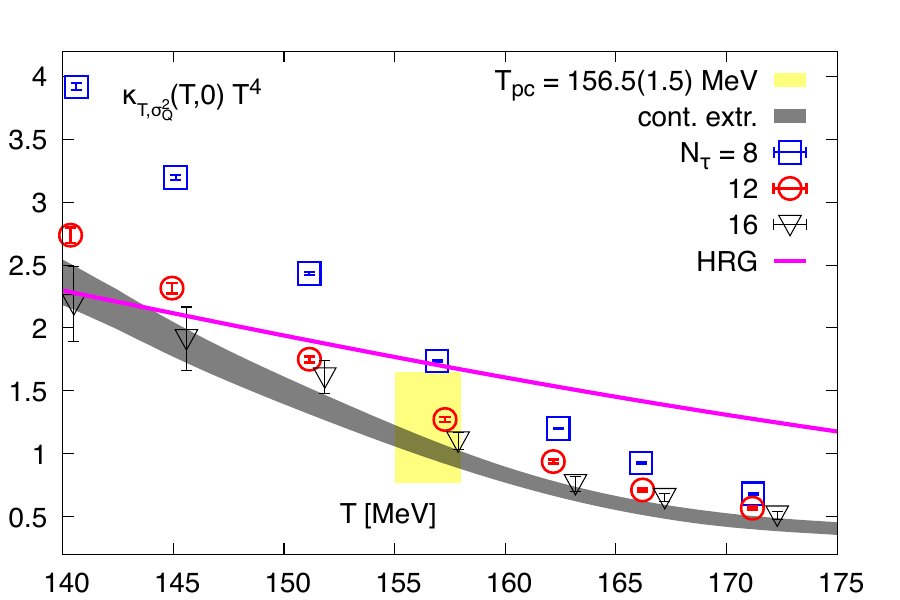}
\includegraphics[width=0.41\linewidth]{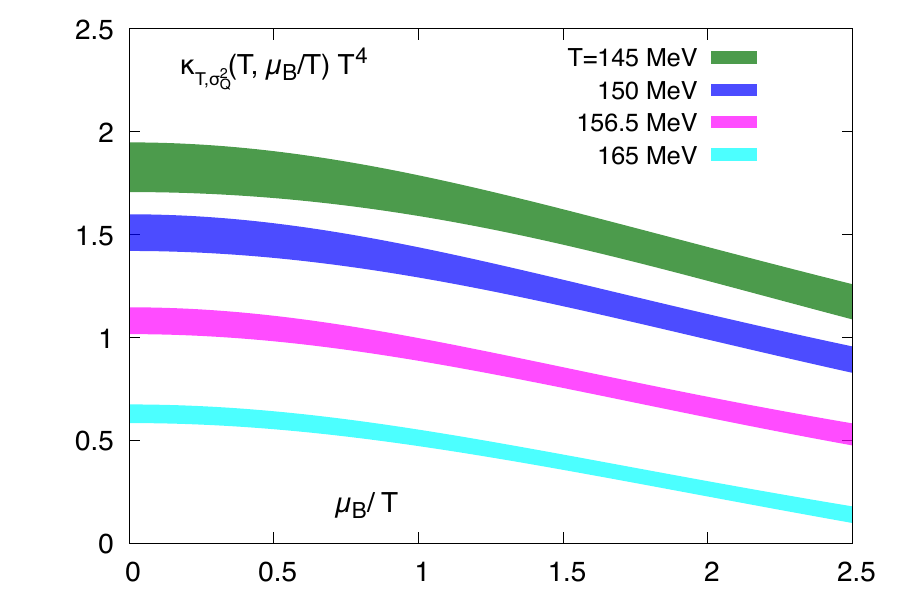}
\caption{Isothermal compressibility of a strangeness-neutral, isospin-symmetric matter for $\hmu_B=0$ ({\it left}) and as function
of $\hmu_B$ at several values of the temperature ({\it right}). 
The grey band in the left hand figure shows the weighted average of continuum limit extrapolations 
obtained with and without including the $N_\tau=8$ data set. Results for $\hmu_B>0$ use Taylor series for the pressure up to sixth order.}
\label{fig:eleckurtosis}
\end{figure*}

The conserved charge cumulants, $\hchi_{ijk}^{BQS}$,
introduced in \eq{eq:suscept}, are
obtained as Taylor series in terms of cumulants calculated 
at vanishing chemical potential, which we denote in the following by $\chi_{ijk}^{BQS}\equiv \hchi_{ijk}^{BQS}(T,\vec{0})$ (i.e. without a ``hat") \cite{Bazavov:2017dus,Bollweg:2022rps}.
The leading-order expansion terms are ${\cal O}(\hmu_B)$ and  
${\cal O}(\hmu_B^0)$ 
for $i+j+k$ odd and even, respectively.
The prefactor $\kappa_{T}^{(0)}$ in \eq{eq:mod-kQ2} may be then be written as
\begin{align}
    \kappa_{T}^{(0)}T^4 = \frac{\hchi_{12}^{BQ}}{\hat{n}_B\hchi_2^Q} &= \frac{\chi_{22}^{BQ}}{\chi_2^B\chi_2^Q}
    \frac{1-\frac{\chi_{121}^{BQS}\chi_{11}^{BS}}{\chi_{22}^{BQ}\chi_2^S}+ {\cal O}(\vec{\hmu}^2)}{1-\frac{(\chi_{11}^{BS})^2}{\chi_2^B\chi_2^S}+ {\cal O}(\vec{\hmu}^2)}\label{eq:k0} \; . 
\end{align}
It gives the dominant contribution to $\kappa_{T,\sigma_Q^2}$. 
It indeed reflects the intuitive understanding that the isothermal compressibility is inversely proportional
to the number of particles in a thermal, strong-interaction medium. In Fig.~\ref{fig:BQ22} we show the
ratio $\chi_{22}^{BQ}/\chi_2^B$. At $\Tpco=(156.5\pm 1.5) ~\rm{MeV}$ \cite{Bazavov:2018mes} its value is close to $0.5$, {\it i.e.}
$(\chi_{22}^{BQ}/\chi_2^B)|_{\Tpco}= 0.41(3)(4)$, the second error reflecting the error on $\Tpco$. 
At  $\Tpco$ the second-order net-electric charge
cumulants in (2+1)-flavor QCD and HRG model calculations differ by less than 10\%. As $\chi_2^Q$ in
non-interacting HRG models is proportional to the total number of charged particles, which represent about
half of the total number of particles at $\Tpco$ (see Table~\ref{tab:kappaHRG}), we see that $\kappa_T^{(0)}$
is a good proxy for the inverse number of hadrons at $\Tpco$, {\it i.e.} at freeze-out in heavy-ion
collisions. Results for $\kappa_T^{(0)}$ obtained in (2+1)-flavor QCD and a non-interacting HRG model 
at $\Tpco$ are given in Table~\ref{tab:kappa_correction}. The difference in both results mainly arises from the 
difference in $\chi_{22}^{BQ}/\chi_2^B$ shown in Fig.~\ref{fig:BQ22}. This difference may be traced back
to the contribution of the doubly charged $\Delta$-resonance to net-electric charge fluctuations. Like all other hadrons the $\Delta$-resonance is, of course, treated as a stable resonance in non-interacting 
HRG models, but is known to be strongly modified in a strong-interaction medium described by QCD \cite{Lo:2017lym}.
We will further elaborate on this in the following.

\begin{table}[htb]
\small
\centering
\begin{tabular}{c|cc|c|c}
\hline\hline
~&$\kappa_T^{(0)} T^4$ & $\delta\kappa_T$ & $\kappa_{T,\sigma_Q^2} T^4$ &$\kappa_{T,\sigma_Q^2}$ [fm$^3$/GeV] \\[4pt]
\hline
QCD  & 1.078(65) & 0.0036(41) & 1.08(10)  & 13.8(1.3) \\
 HRG  & 1.713 & 0.0022 & 1.716 &  21.93\\
\hline\hline
\end{tabular}
\caption{Leading-order isothermal compressibility $\kappa_T^{(0)}$ obtained with \eq{eq:k0}, the correction term $\delta\kappa_T$ , and the 
total isothermal compressibility $\kappa_{T,\sigma_Q^2}$ multiplied with $T^4$ 
evaluated at 
 $\Tpco$ and 
$\muh_B=0$ for (2+1)-flavor QCD and HRG. 
The last column gives $\kappa_{T,\sigma_Q^2}$ at $\Tpco$.
}
\label{tab:kappa_correction}
\end{table}

{\it Isothermal Compressibility in (2+1)-flavor  QCD ---}
We evaluate the isothermal compressibility at fixed
net-electric charge fluctuation in isospin symmetric,
strangeness-neutral matter,
introduced in \eq{eq:mod-kQ2}, using 
a sixth-order Taylor series for $P/T^4$ \cite{Bollweg:2022rps,Bollweg:2022fqq}.
With this we obtain Taylor series for the
first-, second- and third-order 
cumulants appearing in the expression for $\kappa_{T,\sigma_Q^2}$.
Fig.~\ref{fig:eleckurtosis} shows results
for the temperature dependence of $\kappa_{T,\sigma_Q^2} T^4$ 
at $\vec{\mu}=\vec{0}$
and as function of $\hmu_B$ for several values of the temperature. 
 We note that $\kappa_{T,\sigma_Q^2}$
 drops rapidly with 
increasing temperature as well as increasing baryon chemical potential.
With increasing $\hmu_B$ the pseudo-critical 
temperature in (2+1)-flavor QCD 
decreases
\begin{align}
    \Tpc(\hmu_B) = \Tpco\left(1-\kappa_2 \hmu_B^2+{\cal O}(\hmu_B^4)\right)\; .
\label{eq:Tpc}
\end{align}
We use $\Tpco = (156.5 \pm 1.5)\,\text{MeV}$ and
$\kappa_2 = 0.012(4)$, 
from Ref.~\cite{Bazavov:2018mes}, which is consistent within errors with Refs.~\cite{Bonati:2015bha,Borsanyi:2020fev}. 
As a consequence we find that $\kappa_{T,\sigma_Q^2}$, evaluated on the pseudo-critical line, 
has only a weak $\hmu_B$-dependence. 

As pointed out in \eq{eq:kappa-p} for all $T$ the isothermal compressibility of an ideal gas is proportional
to the inverse of the pressure. It thus is reasonable to express $\kappa_{T,\sigma_Q^2}$ in
units of the pressure $P$, which has been calculated in (2+1)-flavor QCD for vanishing \cite{HotQCD:2014kol} as well as non-vanishing \cite{Bollweg:2022fqq} values
of the chemical potentials. On the pseudo-critical line we obtain in next-to-leading order
\begin{align}
    \left. P \kappa_{_{\scriptstyle T,\sigma^2}}\right|_{T_{pc}(\hmu_B)} = 0.80(12) + 0.06(1) \hmu_B^2 +{\cal O}(\hmu_B^4)\; .
    \label{eq:pk}
\end{align}
We thus find that the isothermal compressibility in QCD at the pseudo-critical temperature is close to
that of an ideal gas, {\it i.e} $\kappa_{T,\sigma_Q^2}\simeq 1/P$, as expected when a non-interacting HRG model provides a good description of thermodynamics at the pseudo-critical temperature. 

\begin{figure}
\includegraphics[width=0.90\linewidth]{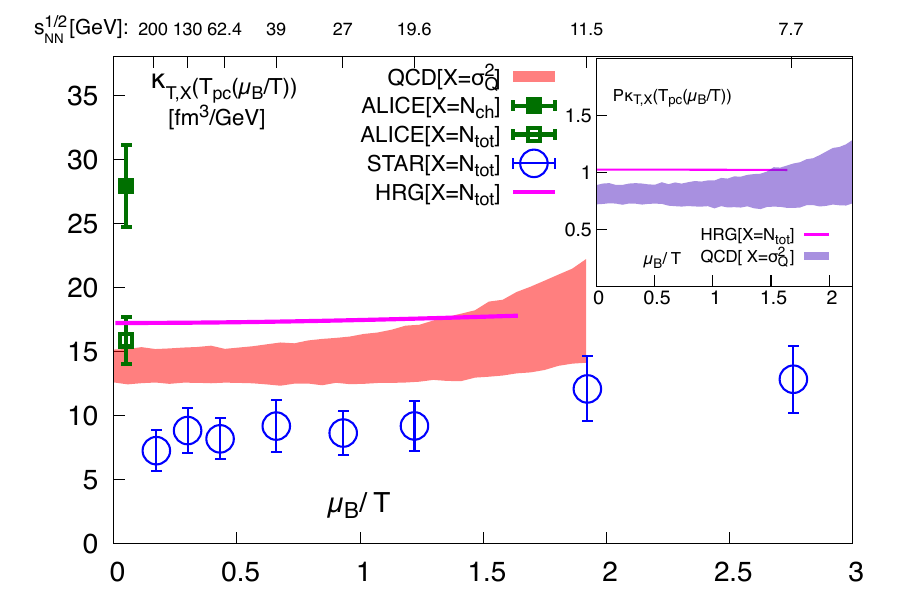}
\caption{Isothermal compressibility $\kappa_{T,\sigma_Q^2}$ (QCD) and $\kappa_{T,\vec{N}}$ (HIC,HRG)
along $\Tpc(\hmu_B)$ (\eq{eq:Tpc}) 
calculated in QCD compared to HIC results. The ${\rm ALICE}[N_{\rm tot}]$ and ${\rm STAR}[N_{\rm tot}]$ data have been obtained by taking into account the contribution of neutral hadrons to the 
total hadron multiplicity as done in \cite{Braun-Munzinger:2014lba}
(see \eq{eq:kinv-approx} and discussion in the text). In the inset we present $P \kappa_{T,X}$ obtained from QCD ($X=\sigma_Q^2$) and  HRG ($X=N_{\rm tot}$) calculations along the line $\Tpc(\hmu_B)$. 
}
\label{fig:isothermal_compressibility_muB}
\end{figure}

In Fig.~\ref{fig:isothermal_compressibility_muB} we show the isothermal compressibility \(\kappa_{T,\sigma_Q^2}\), evaluated along the pseudo-critical line $\Tpc(\hmu_B)$. 
We compare our QCD results for \(\kappa_{T,\sigma_Q^2}\) with those for $\kappa_{T,\vec{N}}$ extracted from heavy-ion collision data 
obtained by the ALICE Collaboration at $\hmu_B=0$ \cite{ALICE:2021hkc} and 
an analysis of data on particle yields and freeze-our radii \cite{STAR:2008med,STAR:2017sal} obtained by the STAR Collaboration.

A similar analysis, taking into account also data from fixed target experiments and using a different approach to determine the thermal component of the 
scaled variance, is given in \cite{Mukherjee:2017elm}. The inset of Fig.~\ref{fig:isothermal_compressibility_muB} shows the dimensionless
product of isothermal compressibility and pressure, $P\kappa_{T,\sigma^2}$ calculated in 
(2+1)-flavor QCD
and $P\kappa_{T,\vec{N}}$ calculated in a non-interacting HRG model. It is evident
that both results are consistent with each other and show
only a mild dependence on $\hmu_B$ as expected from \eq{eq:pk}. They stay close
to the ideal (Boltzmann) gas value, $P\kappa_{T,\vec{N}}=1$. These analyses of HIC data are based on results for charged particle yields and multiplicity distributions
that are used to calculate the isothermal compressibility \(\kappa_{T,\vec{N}}\) at fixed charged-particle numbers using
\eq{eq:kinv-approx} with $n_0= 0$, {\it i.e.} ignoring the contribution of neutral hadrons.
In this case
the ALICE Collaboration finds $\omega_{\rm ch}=1.15(6)$ and $\kappa_{T,\vec{N}} =(27.9\pm 3.18)$~fm$^3$/GeV \cite{ALICE:2021hkc}. Taking into account the contribution of neutral
hadrons by using \eq{eq:kinv-approx} with $\omega_0=\omega_{\rm ch}$ and $N_{\rm tot}=2486(146)$ \cite{Braun-Munzinger:2014lba} one obtains
$\kappa_{T,\vec{N}}=(15.8\pm 1.3)$~fm$^3$/GeV, which is shown in Fig.~\ref{fig:isothermal_compressibility_muB} by an open square. This rescaled ALICE result for $\kappa_{T,\vec{N}}$
compares well with the QCD result obtained at $\vec{\mu}=\vec{0}$,
$\kappa_{T,\sigma_Q^2} = 1.08(10)/ \Tpco^{4} 
=13.8(1.3)~\mathrm{fm}^3/\mathrm{GeV}$ as shown in 
Table~\ref{tab:kappa_correction}.

For $\hmu_B>0$ our calculation of \(\kappa_{T,\sigma_Q^2}\) can be directly confronted with the isothermal compressibility $\kappa_{T,\vec{N}}$ at $\hmu_B>0$ obtained by us 
using data of the STAR Collaboration for charged-particle multiplicities and freeze-out volume at various beam energies \cite{STAR:2008med,STAR:2017sal}. We use results for STAR particle yields ($\pi^\pm,\ K^\pm, p,\ \bar{p}, \Lambda, \bar{\Lambda}$) 
obtained in the beam energy scan at RHIC as compiled in \cite{Chatterjee:2015fua}. With this, we determine the number of charged and total particles in central collisions \cite{Braun-Munzinger:2014lba}. The scaled variances, $\omega_{\rm ch}$, at
various beam energies have been determined by the PHENIX \cite{PHENIX:2008psu} and ALICE collaborations \cite{ALICE:2021hkc}.
The thermal component of the scaled variances is found to be close to unity and compatible with HRG model calculations. 
As shown in Table~\ref{tab:kappa_correction} for $\mu_B=0$ one finds in a non-interacting HRG $\omega_{\rm ch}\simeq \omega_0\simeq \omega_{\rm tot}=1.05$. 
Also for $\hmu_B>0$ we find in HRG model calculations that the scaled variance for charged, neutral as well as total
hadron numbers varies only little on the pseudo-critical line, $\Tpc(\hmu_B)$. We find $\omega_{\rm tot}=1.046$
at $(\Tpc,\mu_B)=(156.5, 0)\ {\rm MeV}$ which decreases to $\omega_{\rm tot}=1.032$
at $(\Tpc,\mu_B)=(144.3,398)\ {\rm MeV}$.
We thus use HRG model results for $\omega_{\rm tot}$ to obtain $\kappa_{T,\vec{N}}$ at various RHIC beam energies
on the pseudo-critical line $\Tpc(\hmu_B)$.
The results are also shown in Fig.~\ref{fig:isothermal_compressibility_muB}. We note that $\kappa_{T,\vec{N}}$, extracted
from the STAR data, rises slowly with increasing $\hmu_B$. This, however, is solely driven by the decrease of $N_{\rm tot}$
with decreasing $\sqrt{s_{_{NN}}}$ and not due to an increase in fluctuations of the particle multiplicities. We thus
have no evidence for reaching the vicinity of a critical point where these fluctuations and as such also $\kappa_{T,\vec{N}}$ 
should diverge.

{\it Summary ---}
We introduced and examined the properties of a generalized isothermal compressibility which is suitable for the
analysis of compressibility of matter in which the number of particles is not a conserved quantity. We calculated
the compressibility of strong-interaction matter keeping temperature and the fluctuations of net-electric charge
fixed, which in a non-interacting HRG is a good proxy for total charged particle numbers. 
We found that this generalized  compressibility varies little on the pseudo-critical line and stays close to the 
value $\kappa_{T,\sigma_Q^2}=13.8(1.3)$~fm$^3$/GeV obtained at $(\Tpc,\mu_B)=(156.5,0)\ {\rm MeV}$. This finding
is consistent with the rescaled result of the ALICE Collaboration, where we replaced the number of charged hadrons
by the total number of hadrons at freeze-out. It suggests that the isothermal compressibility of strong-interaction matter 
along the pseudo-critical transition line, 
or at the time of hadronization in HIC, stays close to that of an ideal (Boltzmann)  gas. 


{\it Acknowledgements ---}
We thank Arghya Chatterjee, Sandeep Chatterjee and Krzysztof Redlich for very helpful discussions. We acknowledge the HotQCD Collaboration for providing the gauge configurations. 
JG would also like to thank Dibyendu Bala for helpful discussions.
This work was supported by The Deutsche Forschungsgemeinschaft (DFG, German Research Foundation) - Project numbers 315477598-TRR 211 and 460248186 (PUNCH4NFDI); by the U.S. Department of Energy, Office of Science, Office of Nuclear Physics through Contract No. DE-SC0012704,
the Funding Opportunity Announcement Scientific Discovery through
Advanced Computing: High Energy Physics, LAB 22-2580, and by the National Science Foundation under Grants
No. PHY20-13064 and PHY23-1057.
We use the AnalysisToolbox \cite{Clarke:2023sfy,githubAnalysistoolbox} for data analysis.

\bibliography{bibliography}

\end{document}